\documentstyle[11pt,aaspp4]{article}

\def\etal   {{et~al.\ }}

\lefthead{Udomprasert, Taylor, Pearson \& Roberts }
\righthead{RM structure of OQ~172}

\begin{document}

\phantom{~}
\vspace{2cm}

\title{Evidence for Ordered Magnetic Fields \\ in the Quasar Environment}

\author{P.S.~Udomprasert\altaffilmark{1}, G. B. Taylor\altaffilmark{2}, T.J. Pearson\altaffilmark{3} \& D.H. Roberts\altaffilmark{4}}

\altaffiltext{1}{Dept. of Astrophysics, Princeton University, Princeton NJ, 08544; \\ psu@astro.princeton.edu}
\altaffiltext{2}{NRAO, Box 0, Socorro, NM, 87801; gtaylor@nrao.edu}
\altaffiltext{3}{California Institute of Technology, 105-24, Pasadena CA, 91125; 
tjp@astro.caltech.edu}
\altaffiltext{4}{Dept. of Physics, MS-057, Brandeis University,
Waltham MA, 02254; \\ dhr@vlbi.astro.brandeis.edu }

\vspace{1cm}
\hfil{~~~~~~~~~~Astrophysical Journal Letters, in press}\hfil
\vspace{1cm}

\begin{abstract}

At a distance of 20 pc from the purported supermassive black hole
powering quasars, temperatures and densities are inferred from optical
observations (\cite{ost89}) to be $\sim10^4$ K and $\sim10^4$
cm$^{-3}$.  Here we present Very Long Baseline Interferometry radio
observations revealing organized magnetic fields on the parsec scale
in the hot plasma surrounding the quasar OQ~172
(1442+101).  These magnetic fields rotate the plane of polarization of
the radio emission coming from the core and inner jet of the quasar.
The derived rotation measure (RM) is 40,000 rad m$^{-2}$ in the rest
frame of the quasar.  Only 10 mas (a projected distance of 68 pc) from
the nucleus the jet absolute values of RM fall to less than 100 rad
m$^{-2}$.

\end{abstract}

 
\vspace{1cm}

\section{Introduction}

While typical extragalactic radio sources display Faraday RMs of
1--100 rad m$^{-2}$ at arcsecond resolution, OQ~172 is one of only
$\sim$20 sources found to have RMs exceeding 1000 rad m$^{-2}$
(\cite{tay92,ino95,car96}).  About half of these are Compact Steep
Spectrum (CSS) sources.  Those of larger angular extent, for example,
Hydra~A, Cygnus~A, and 3C~295, have been shown by VLA observations to
possess ordered RM structure on scales of 1--100 kpc
(\cite{dre87,per91,tay93}).  The origin of these high RMs is a
magnetized X-ray emitting cluster gas with field strengths of $\sim$
10 $\mu$G (\cite{tay94}).  In the case of the high RM CSS sources,
observations at a resolution of 1 arcsecond or more were insufficient
to measure the scale of the magnetic fields and thereby determine
their origin.  Our multi-frequency VLBA polarimetric observations of
the CSS quasar OQ~172 in the 6 and 18 cm bands agree with the high RMs
seen at lower resolution (M. Inoue, private communication) and further
reveal that the high RMs are associated with the nuclear component.
Such differences in the RM structure on these small scales rule out a
cluster magnetic field Faraday screen.  We suggest that the high RMs
seen in OQ~172, and perhaps in other CSS quasars as well, originate in
the nuclear environment.

\cite{osm94} have measured the redshift of OQ~172
to be z=3.52.  Assuming {\nobreak$H_0$ = 50 km s$^{-1} $Mpc$^{-1}$} and
$q$=0.5, 1 mas corresponds to 6.81 pc.

\section{Observations and Results}

The observations, performed on 1995 June 4, were carried out at four
frequencies in the 6 and 20 cm bands using the 10 element VLBA and one
element of the VLA (see Table 1. for details of the parameters
used).  Observations of the strong calibrator OQ~208 were employed for the
polarization calibration following the standard procedure
(\cite{cot93,rob94}).  A short segment of data from the strongly
polarized source 3C~345 (\cite{bro94}) was used for calibration of the
polarization angles.  A single correction was applied for all 6 cm
frequency observations of OQ~172.  By applying a constant correction,
we preserve the relative angles among the different frequencies, which
are of prime importance to this experiment.
  
The four frequencies within the 6 cm band were combined to create a
total intensity image of OQ~172 (Fig.~1).  Based on compactness,
relative strength, and a flat spectrum (see below), we identify the
strong north-easternmost component in Fig.~1 as the core.  At this
resolution, OQ~172 shows the typical core and jet structure of a CSS
quasar (\cite{dal95}), but it is slightly unusual in that the jet
turns through almost $180^\circ$.  The jet emission extends from the
central core in a WSW direction and almost immediately bends
southward.  About 20 mas south of the core, it turns again, about
$90^\circ$, and extends to the east.  No extended emission has been
detected on scales greater than 100 mas (\cite{spe89}).  A spectral
index image was made from the total intensity images at 6 cm and 20 cm
(Fig.~2).  The core shows the flattest spectrum ($\alpha = -0.4$).
However, because the core and the inner jet are unresolved it is
likely that the core component has an even flatter spectrum.  The
spectral index of the jet is very steep (typically, $\alpha = -1.2$).
This is steeper than the expected value of $\alpha = -0.7$ and is
probably caused by a loss of flux at 6 cm due to a lack of short
spacings.

\def\dg{$^{\circ}$}
\begin{center}

TABLE 1 \\
\smallskip
VLBA O{\sc bservational} P{\sc arameters}
\smallskip
 
\begin{tabular}{l r r r r r r r r}
\hline
\hline
Source & Frequency & BW & $\Delta\nu$ & $\Delta$t & Scan & Time \\
 (1) & (2) & (3) & (4) & (5) & (6) & (7) \\
\hline
\noalign{\vskip2pt}
OQ~172 & 1.506, 1.514, 1.664, 1.672  & 7 & 500 & 4 & 8.5 & 94 \\
        & 4.616, 4.654, 4.854, 5.096 & 7 & 500 & 4 & 8.5 & 85 \\
OQ~208  & 1.506, 1.514, 1.664, 1.672 & 7 & 500 & 4 & 4 & 40 \\ 
        & 4.616, 4.654, 4.854, 5.096  & 7 & 500 & 4 & 4 & 40 \\ 
3C~345 & 1.506, 1.514, 1.664, 1.672  & 7 & 500 & 4 & 4 & 4 \\
       & 4.616, 4.654, 4.854, 5.096 & 7 & 500 & 4 & 4 & 4 \\
\hline
\end{tabular}
\end{center}
\smallskip
\begin{center}
{\sc Notes to Table 1}
\end{center}
\begin{flushleft}
Col.(1).---Source name.
Col.(2).---Observing frequencies in GHz.
Col.(3).---Total spanned bandwidth in MHz.
Col.(4).---Channel width output from correlator in kHz.
Col.(5).---Integration time output from correlator in seconds.
Col.(6).---Length of each scan in minutes.
Col.(7).---Total integration time on source in minutes.
\bigskip
\end{flushleft}

Our 20 cm total intensity image of OQ~172 compares well with one
made at 1.66 GHz by \cite{dal95} using a global
array.  A 6 cm image produced by \cite{gur94}, however, is
inconsistent with any of the images produced by us or by Dallacasa
et~al..  The Gurvits \etal image is now thought to suffer from
calibration errors.

The RM structure of OQ~172 was determined by comparing the four
position angle (PA) images in the 6 cm band and determining the change
in PA with respect to wavelength squared for each pixel.  The
resulting RM image, shown in Fig.~3, reveals a distinct difference in RM
between the core and the southern jet.  Typical RM values at the core
are extremely high, ranging from 1000 to 2000 rad m$^{-2}$, with a
gradient of about $-$445 rad m$^{-2}$ mas$^{-1}$
toward the southwestern jet, whereas values in the southern jet area
range from $-$200 to $-$60 rad m$^{-2}$.  Fig.~4 shows sample fits to
the position angle vs. wavelength squared for a group of pixels in the
region of the core.  The fits in the core agree well with a $\lambda^2$ law in
the core.  In the jet, the lower rotation measures yield a much
smaller change in position angle, approaching the minimum RM
detectable over this frequency range.  At 20 cm we found the inner
region to be unpolarized, consistent with depolarization by the 
large RM gradients seen at 6 cm.  The southern
and eastern portions of the jet were found to be up to 15\% polarized,
with RMs of $-80 \pm 20$ rad m$^{-2}$.

The most striking feature of this high resolution RM image (Fig.~3) is
the observed non-uniformity between the inner and outer jet.  The
lower RMs observed in the outer, southern jet are comparable to RM
values obtained from single dish measurements of nearby objects (\cite{sim81})
and can thus be explained as being
Galactic in origin.  However, something particular to the nuclear
environment must be producing the high RMs in the inner
core and jet of the quasar.  In this region the RMs in the rest frame
of OQ~172 are 20,000 to 40,000 rad m$^{-2}$, values which, to our
knowledge, are the highest RMs yet observed in any extragalactic
source.  Since the scale length for these RMs is so small ($\sim$70
pc), they cannot be explained by the cluster-scale magnetic fields
that cause the high RMs in the larger radio galaxies such as Hydra~A
and Cygnus~A.  The high RMs observed in OQ~172 thus result from an
effect confined to the nuclear region.  Here we discuss some of the
possible candidates.

\section{Discussion}

We first consider internal Faraday rotation as the cause of the high
observed RMs.  In internal Faraday rotation, thermal material is mixed
in with the synchrotron emitting plasma.  For a slab model in which
the density and magnetic field are homogeneous throughout the
source (\cite{bur66}), the position angle of the polarized flux density,
$\Psi$, follows a wavelength squared law, but only between 0 and
90$^\circ$ with discontinuous jumps in $\Psi$.  There is also
considerable depolarization, with nulls at the discontinuous changes
in polarization angle.  Fig.~4 shows that for various pixels, $\Psi$
turns through almost 90$^\circ$ without a discontinuity or marked
change in the fractional polarization, and we thus rule out internal
Faraday rotation.

We now consider a magnetized plasma in the nuclear region, but
external to the radio source, as the cause of the RMs observed in the
core.  For such an external Faraday screen the
rotation measure is given by (\cite{spi78})
\begin{equation}
\label{RM}
RM = 812\int\limits_0^L n_{\rm e} B_{\|} {\rm d}l ~{\rm radians~m}^{-2}~,
\end{equation}
\noindent 
where $n_e$ is the electron density in cm$^{-3}$, $B_{\|}$
is in mGauss, and $l$ is in pc.  On average the magnetic field, $B$,
will be larger than the parallel component, $B_{\|}$, by $\sqrt{3}$.
If the magnetic field is tangled, then it will be larger by 
a factor $N^{1/2}$ where $N$ is the number of cells along the 
line of sight.

Two regions containing thermal gas which could act as a Faraday screen
if magnetized are the broad line region (BLR) and the narrow line
region (NLR) associated with active galactic nuclei.  The BLR is
thought to have a size of $\sim$ 0.1 pc and magnetic field strengths
of perhaps 1 G (\cite{ree87}).  However the region of high RM in OQ~172
extends over 3 mas, or a projected size of 20 pc.  Given the likely 
inclination of the jet close to the line-of-sight, this region could
be considerably larger.  Thus the BLR is unlikely to produce
coherent RMs on the observed scale.  The NLR is thought to cover a
region of 100-1000 pc albeit with a small volume filling factor
($\epsilon \sim$ 10$^{-3}$ [\cite{ost89}]).  In order to produce a coherent RM on
scales of tens of pc, the covering factor of the NLR must be close to
unity, and the free-free opacity must be sufficiently low that the
radio source is not extinguished at cm wavelengths.  Free-free
opacities for the NLR, with the standard estimates for density of
10$^4$ cm$^{-3}$ and temperature of 10$^4$ K, are much less than unity
at cm wavelengths (\cite{ulv81}).  The covering factor could be close to
unity if the NLR clouds form preferentially in the vicinity of the
radio jet (\cite{nor84}).  If the magnetic field is in pressure equilibrium
with the NLR then its field strength would be 0.6 mG.  A RM of 40,000
rad m$^{-2}$ could be produced in a pathlength of 0.01 pc.  
Such a short pathlength is unlikely to produce
a coherent RM since variations across the source would result in
depolarization.  It is more likely that the magnetic fields are not in
equipartition -- if distributed over 20 pc the field strengths in the
NLR are $\sim$0.5 $\mu$Gauss.

Alternatively, the NLR clouds could be thermally confined by a more
rarefied intercloud region at higher temperature.  \cite{ode89} took
a density of 10 cm$^{-3}$, temperature 10$^7$ K, and size of 100 pc
for this intercloud material and assuming an equipartition magnetic
field strength of 1 mG, predicted a RM of 2 $\times$ 10$^5$ rad
m$^{-2}$ toward the cores of quasars.  A tangled magnetic field, or a
shorter path length through this intercloud region could produce the
RM observed for the core of OQ~172.  O'Dea observed much lower
$|$RM$|$s of $\sim$200 rad m$^{-2}$ for a sample of 15 core dominated
quasars and BL Lac objects.  His measurements, however, were based on
VLA observations at $\sim$150 mas resolution.  At this comparatively
low resolution his observations may not reflect the true RM of the
core component.


Finally we ask ourselves what is special about OQ~172?  It is
possible that we are seeing it along some favorable line-of-sight, or
that the magnetic fields of the NLR are enhanced in this source.  
Few sources, however, have been studied in this way.  In fact the low
polarization seen in many quasar cores at 6 cm (\cite{caw93}) may be the
result of depolarization from a Faraday screen.  Just recently, 
multi-frequency VLBI polarimetry of the quasar 3C~138 has revealed a RM of 
$-1780 \pm 50$ within 22 pc from the nucleus while the jets further out exhibit low RMs (\cite{cot96}).  OQ~172 is also
similar to other quasars in that, once corrected for rotation measure,
the polarized emission appears to follow the curve of the jet (\cite{rob89,bro94}).  RM
studies performed to date at arcsecond resolution are in almost all
cases dominated by the polarized flux of jets (\cite{rob89}).  The
jets will have lower RMs if they are 
outside the region of high magnetic fields.
The ``anomalous'' high RM CSS sources discovered in surveys at low
resolution may be unusual only in that their cores dominate the
integrated polarized flux density and are therefore detectable in the
integrated RM.  High resolution RM measurements of additional sources, especially
``typical'' quasars are needed.  This technique may lead us to a
better understanding of the nuclear environment in AGNs.


\acknowledgments

One of us (PSU) acknowledges support from a summer research
position at NRAO.  We acknowledge use of the VLA and VLBA of The
National Radio Astronomy Observatory which is operated by Associated
Universities, Inc., under cooperative agreement with the National
Science Foundation.  This research has made use of the NASA/IPAC
Extragalactic Database (NED) which is operated by the Jet Propulsion
Laboratory, Caltech, under contract with NASA.  This research also
made use of data from the University of Michigan Radio Astronomy
Observatory which is supported by the National Science Foundation and
by funds from the University of Michigan.  TJP acknowledges support
by the NSF under grant AST-9420018.  DHR acknowledges support
by the NSF under grants AST-91022282 and AST-95-29228.

\clearpage

\begin{figure}
\vspace{15.5cm}
\includegraphics{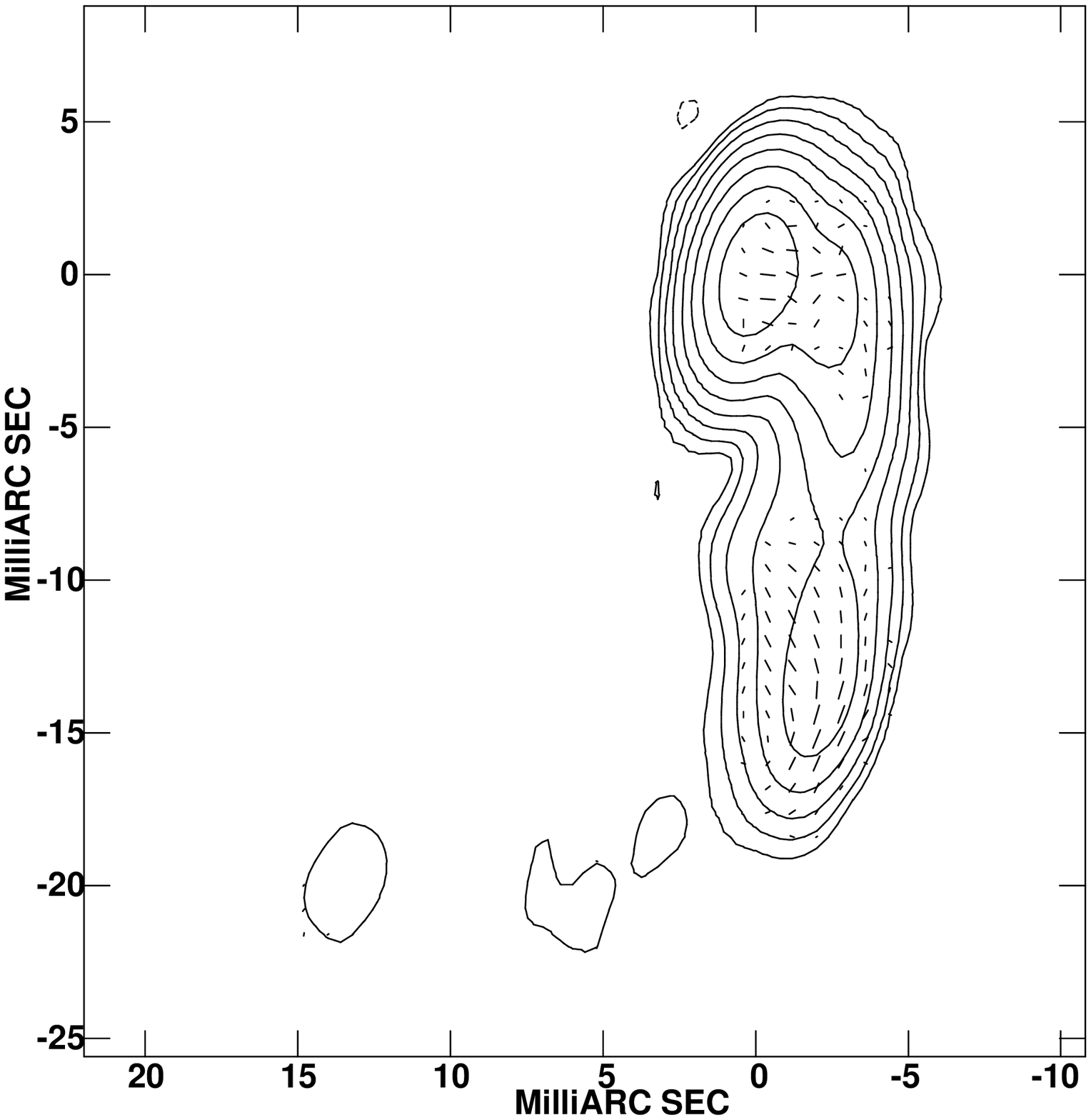}
\figcaption{A contour image of OQ~172 at 6 cm with a resolution of
4.2 $\times$ 1.7 mas in position} 
\noindent
angle $-$15\arcdeg.  Contour levels are at 
$-$1.7, 1.7, 3.4, 6.8, 14, 27, 54, 109 and 218 mJy/beam with negative contours shown as dashed lines.
Positions are given in milliarcseconds
relative to the strongest component, located at RA(J2000) 14:45:16.465
and Dec(J2000) 09:58:36.072 (\cite{joh95}).  This naturally
weighted image has a dynamic range of 1700.  Overlaid on the contours is the 
RM-corrected projected magnetic field of OQ~172.  A vector length
of 1 mas corresponds to a polarized intensity of 10.4 mJy/beam at
6 cm.  The error in the projected magnetic field direction ranges 
from 10 -- 20$^\circ$.
\end{figure}
\clearpage

\begin{figure}
\vspace{15.5cm}
\includegraphics{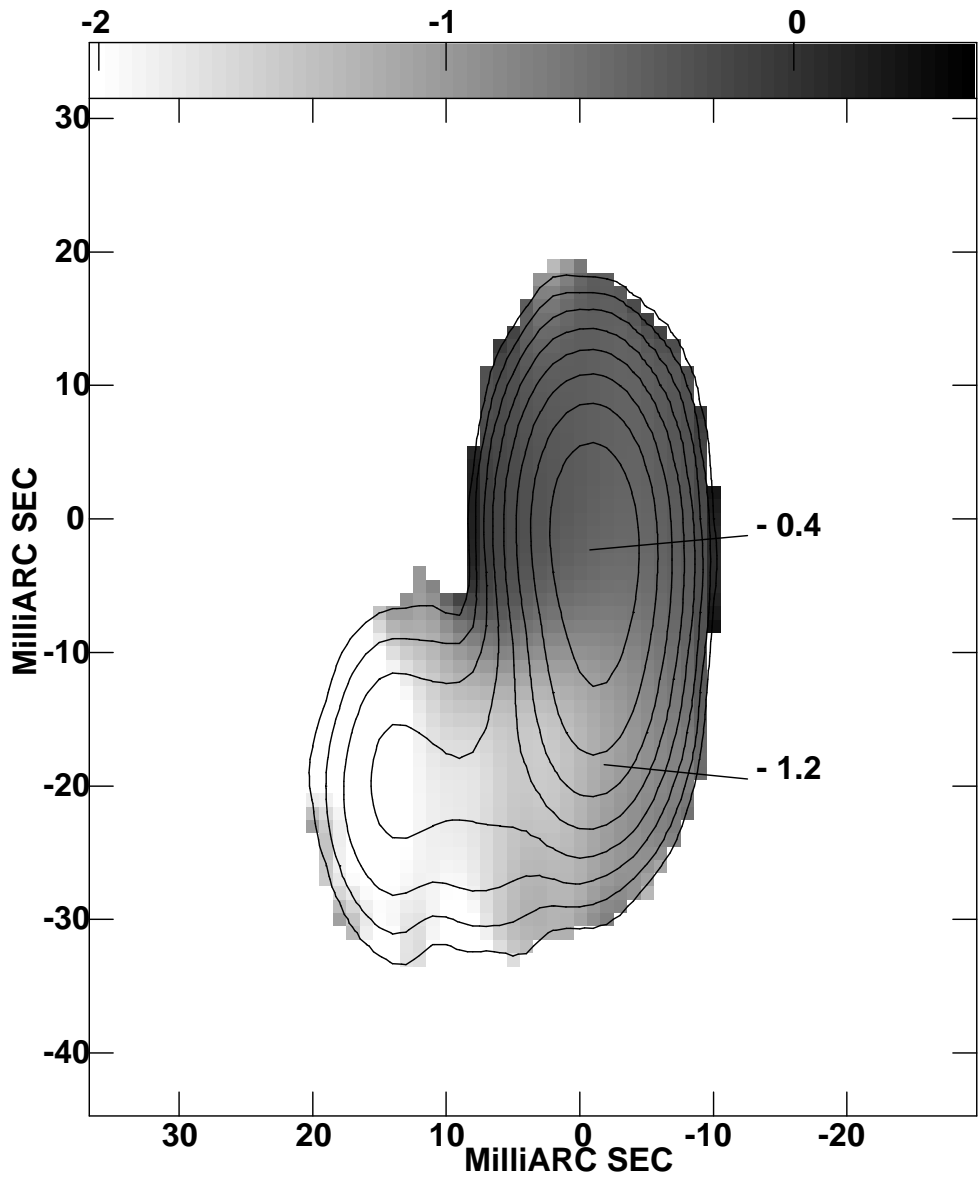}
\figcaption{The spectral index of OQ~172 between 6 cm and 20 cm.  Since the two frequencies yielded images of different
resolutions, the 6 cm image had to be remade with a taper to match the
resolution of the 20 cm image.  The
restoring beam of the total intensity images has dimensions 13
$\times$ 6 mas in position angle 0\arcdeg.  The grey scale range is
from $-$2 to 0.5.  Contours from the 20 cm image are overlaid at $-$5.4,
5.4, 11, 22, 43, 86, 173, 346, and 691  mJy/beam.}
\end{figure}
\clearpage

\begin{figure}
\vspace{18.5cm}
\includegraphics{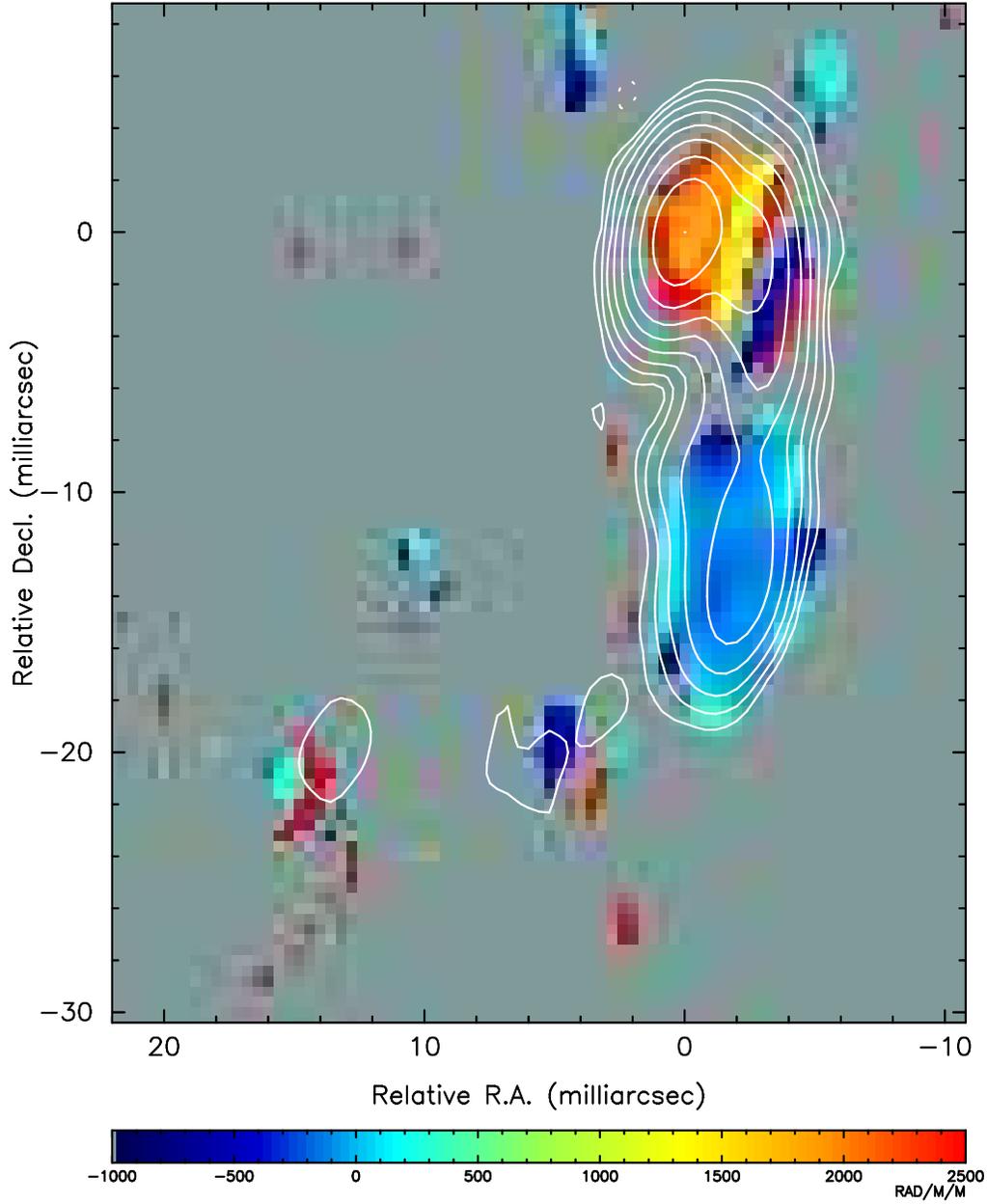}
\figcaption{False color rotation measure image of OQ~172 at 4 $\times$ 2 mas resolution.  
A pixel was blanked if the error in PA exceeded $20^\circ$ for
any particular frequency.  No corrections have been made for the
redshift of the observed emission, so if the RMs are being produced in
the vicinity of the source, the values in the rest frame of OQ~172 are
larger by a factor of $(1+z)^2$, or 20. The
colorbar range is from $-$1000 to 2500 rad m$^{-2}$.  Contours are as
in Figure 1.}
\end{figure}
\clearpage

\begin{figure}
\vspace{15.5cm}
\includegraphics{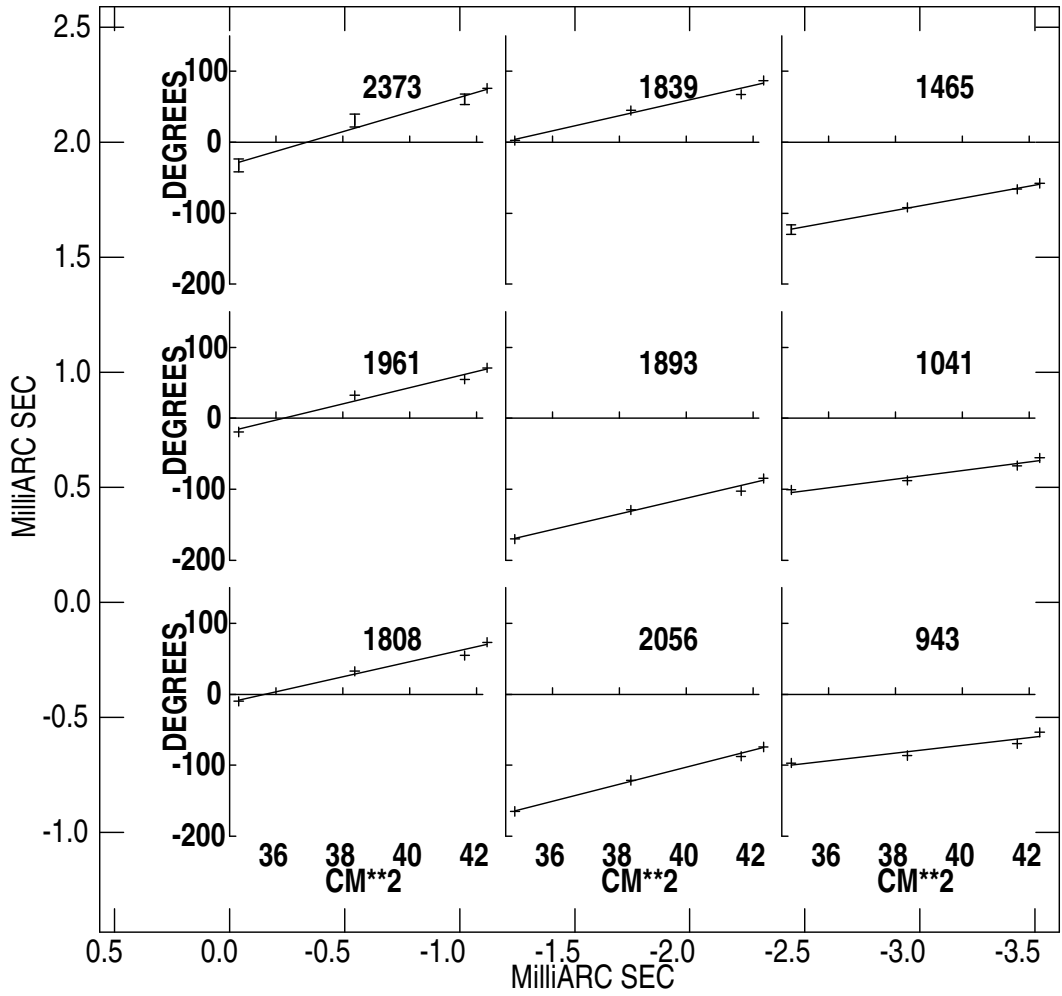}
\figcaption{Sample fits to the polarization angle versus wavelength
squared in the core region of OQ~172 
for the 6 cm frequencies (4612, 4650, 4850 and 5092 MHz).  The fits
are plotted every 1.2 mas. 
Fairly good agreement with a $\lambda^2$ law is found.}
\end{figure}

\clearpage
\end{document}